\documentclass[pre,twocolumn,usletter,showpacs,superscriptaddress]{revtex4}

\usepackage{amsmath}
\usepackage{amsfonts}
\usepackage{amssymb}

\usepackage [ansinew] {inputenc}
\usepackage[dvips]{graphicx}
\usepackage{natbib}

\begin{document}
\title{Origin of Complexity and Conditional Predictability \\
in Cellular Automata}

\author{Vladimir Garc\'{\i}a-Morales}
\email{vmorales@ph.tum.de}
\affiliation{Institute for Advanced Study - Technische Universit\"{a}t M\"{u}nchen, Lichtenbergstr. 2a, D-85748 Garching, Germany}
\affiliation{Nonequilibrium Chemical Physics - Physics Department - Technische Universit\"{a}t M\"{u}nchen, James-Franck-Str. 1, D-85748 Garching, Germany}

\begin{abstract}
\noindent A simple mechanism for the emergence of complexity in cellular automata out of predictable dynamics is described. This leads to unfold the concept of conditional predictability for systems whose trajectory can only be piecewise known. The mechanism is used to construct a cellular automaton model for discrete chimera-like states, where synchrony and incoherence in an ensemble of identical oscillators coexist. The incoherent region is shown to have a periodicity that is three orders of magnitude longer than the period of the synchronous oscillation. 
 \end{abstract}
 \maketitle \pagebreak

\title{} 

\section{Introduction}

Complexity involves the interaction of many highly correlated individual units that lead to nontrivial emergent behavior \cite{Prigogine, Haken, Mainzer, Holland, Chua, Kuramoto2, Gellmann, Kauffman, Wolfram1}. Complex patterns are ubiquitously found in nature. Mollusc seashells, snowflakes, DNA strands just provide a few examples. In Physics, self-organized criticality \cite{Bak} is a robust mechanism that  leads to complexity out of very simple rules. To date there is, however, no known set of general features that warrant that a system will have this property. 

\begin{figure}
\includegraphics[width=0.45\textwidth]{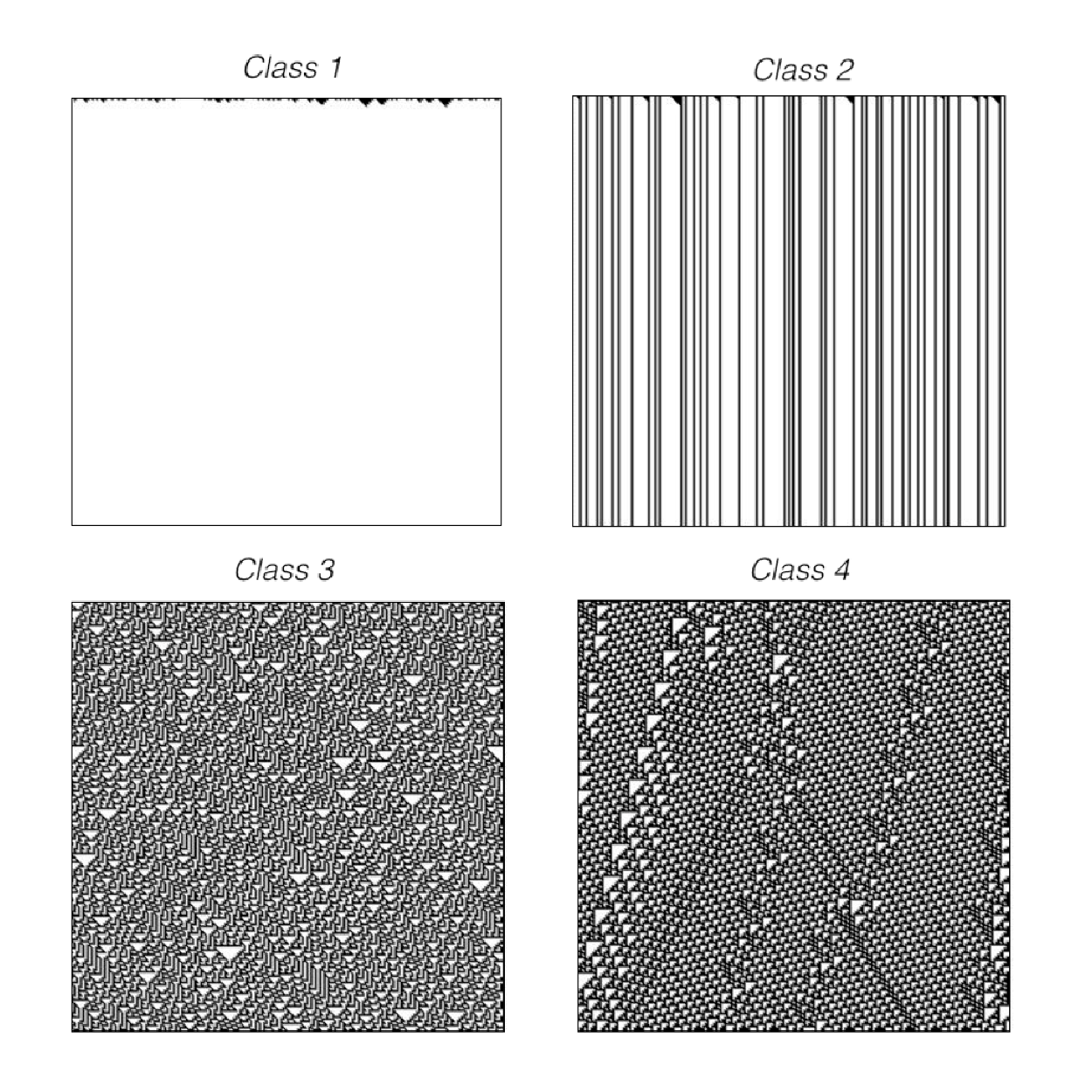}
\begin{center}
\caption{Spatiotemporal evolution of one-dimensional cellular automata rules representative of each of the Wolfram classes of increasing complexity. Wolfram's rules 254 (Class 1), 232 (Class 2), 30 (Class 3) and 110 (Class 4). Time flows from top to bottom.} \label{Wolfy}
\end{center} 
\end{figure}

Another approach is provided by cellular automata (CAs) \cite{Wolfram1, Neumann, Codd, morales1, morales2, Wolfram2, Wolfram3, Wolfram4, Wolfram5, Wolfram6}. These systems have a finite number of states $p$ and evolve on a discrete space-time by means of homogeneous local rules which depend on the dynamical states of an spatial range $\rho$ of neighboring locations. In spite of their simplicity, CAs are able to capture many patterns observed in nature and play often analogous roles in discrete systems to partial differential equations in continuum ones. From computer experiments, Wolfram classified CAs into four classes of increasing complexity \cite{Wolfram2} (see Fig. \ref{Wolfy}): For a random initial condition a CA evolves into a single homogeneous state (Class 1), a set of separated simple stable or periodic structures (Class 2), a chaotic, aperiodic or nested pattern (Class 3) or complex, localized structures, some times long-lived (Class 4). For the latter class \emph{no shortcut is possible in general to find the trajectory, neither qualitatively nor quantitatively} \cite{Wolfram1, Wolfram5}. Wolfram's classification is purely based on how the spatiotemporal evolution of CA rules look like \cite{Wolfram1}. 

Finding Class 4 CA rules requires huge computational explorations in rule space \cite{Wolfram1}: no systematic approach exists to obtain them. Since the total number of CA rules increases dramatically with the number of discrete dynamical states $p$ and neighborhood range $\rho$ (as $p^{p^{\rho}}$) such brute force approach is useless for $p$ or $\rho$ sufficiently large. This article gives a simple prescription to directly find Class 4 CA rules. The approach is valid for any values of $p$ and $\rho$. It is based on 
a symmetry breaking process that certain complex (yet predictable) Class 3 CA rules possess: the invariance under addition modulo $p$. The mathematical tool that is used thorough is $\mathcal{B}$-calculus \cite{morales1, morales2} which provides a unified framework to describe rule-based dynamical systems, as cellular automata and substitution systems.

The outline of this paper is as follows. In Section \ref{invariance} we describe the invariance under addition modulo $p$, an essential symmetry that is broken by Class 4 CA. In Section \ref{breaking} we describe a simple mechanism of symmetry breaking that leads to Class 4 complexity. In Section \ref{conditional} this mechanism is illustrated to introduce the concept of conditional predictability in CA: rules are invented that lead to predictable, nested, Class 3 behavior in short time and space scales but Class 4 behavior in larger scales. In Section \ref{chimeras}, the theory is applied to construct a model for discrete chimera-like states.

\section{Invariance under addition modulo $p$}
\label{invariance}
 
A CA dynamics on a ring with total number of sites $N_{s}$ is considered.  The input initial condition is a vector $(x_{0}^{1},...,x_{0}^{N_{s}})$. Each $x^{i}_{0}$ is an integer $\in [0,p-1]$ where superindex $i$ specifies a position on the ring ($i \in [1,N_{s}]$). At each $t$ the vector $(x_{t}^{1},...,x_{t}^{N_{s}})$ specifies the state of the CA. Inputs and outputs from the CA rule are integers on the interval $[0,p-1]$. Periodic boundary conditions are considered ($x_{t}^{N_{s}+1}=x_{t}^{1}$ and $x_{t}^{0}=x_{t}^{N_{s}}$). The output of totalistic CA rules at each site $i$ is a function of the sum over previous site values on a neighborhood of range $\rho=l+r+1$ as $x_{t+1}^{i}=f\left(\sum_{k=-r}^{l}x_{t}^{i+k}\right)$, where $l$ and $r$ are the number of sites to the left and to the right of site $i$, respectively. We take the convention that $i$ is more positive to the left. \emph{Invariance under addition modulo $p$ means $f\left(mp+y\right)=f\left(y\right)$ for $m$ a natural number and $y \equiv \sum_{k=-r}^{l}x_{t}^{i+k}$}. We now prove that invariance under addition modulo $p$ implies also \emph{discrete scale invariance} \cite{sornette}: We have $f(\lambda y)=f(y)$ for a countable infinite set of $\lambda$ values. We write $\lambda$ in base $p$ as $\lambda=\lambda_0+\sum_{n=1}^{N}\lambda_n p^n$, with all $\lambda_n$ integers $\in [0,p-1]$. Then, all dilatations with $\lambda_0=1$ are $\lambda=1+m'p$, with $m'=\sum_{n=1}^{N}\lambda_n p^{n-1}$ integer and, by using the invariance modulo $p$ and that $m \equiv m'y$ is integer, $f(\lambda y)=f((1+m'p) y)=f(y+m'py)=f(y+mp)=f(y)$. This property qualitatively encodes the \emph{lacunarity} of a fractal structure \cite{sornette, Mandelbrot}. Any rule of the form 
 \begin{eqnarray}
x_{t+1}^{i}=\left(\sum_{k=-r}^{l}x_{t}^{i+k}\right)^{h} \qquad \mod p \label{wunder}
\end{eqnarray} 
is invariant under addition modulo $p$ for a non-negative integer $h$. This can be easily proved as follows
\begin{eqnarray}
&&f\left(mp+\sum_{k=-r}^{l}x_{t}^{i+k}\right)=\left(mp+\sum_{k=-r}^{l}x_{t}^{i+k}\right)^{h} \mod p \nonumber \\
&=& \sum_{j=0}^{h}\binom{h}{j}\left(\sum_{k=-r}^{l}x_{t}^{i+k}\right)^{h-j}\left(mp\right)^{j} \mod p \nonumber \\
&=& \left(\sum_{k=-r}^{l}x_{t}^{i+k}\right)^{h} \mod p = f\left(\sum_{k=-r}^{l}x_{t}^{i+k}\right) \nonumber
\end{eqnarray}
where in getting to the next-to-the-last equality, it has been used that all terms in the sum for which $j \ne 0$ are proportional to $p$ and thus cancel modulo $p$. It is important to remark that the operation $\mod p$ only warrants that the result is an integer $\in [0,p-1]$. This, however, does not mean that an expression $\mod p$ is invariant under addition modulo $p$. For example $f\left(x_{t}^{i}+x_{t}^{i-1}\right)=-\left(x_{t}^{i}+x_{t}^{i-1}\right)^2/2+3\left(x_{t}^{i}+x_{t}^{i-1}\right)/2 \mod 2$, where $x_{t}^{i}$ and $x_{t}^{i-1}$ take only integer values '0' or '1', is not invariant under addition modulo 2: configurations with neighborhood values $(x_{t}^{i},x_{t}^{i-1})=(0,0)$ and  $(x_{t}^{i},x_{t}^{i-1})=(1,1)$ output different values '0' and '1', respectively.

\begin{figure} 
\includegraphics[width=0.3\textwidth, angle=270]{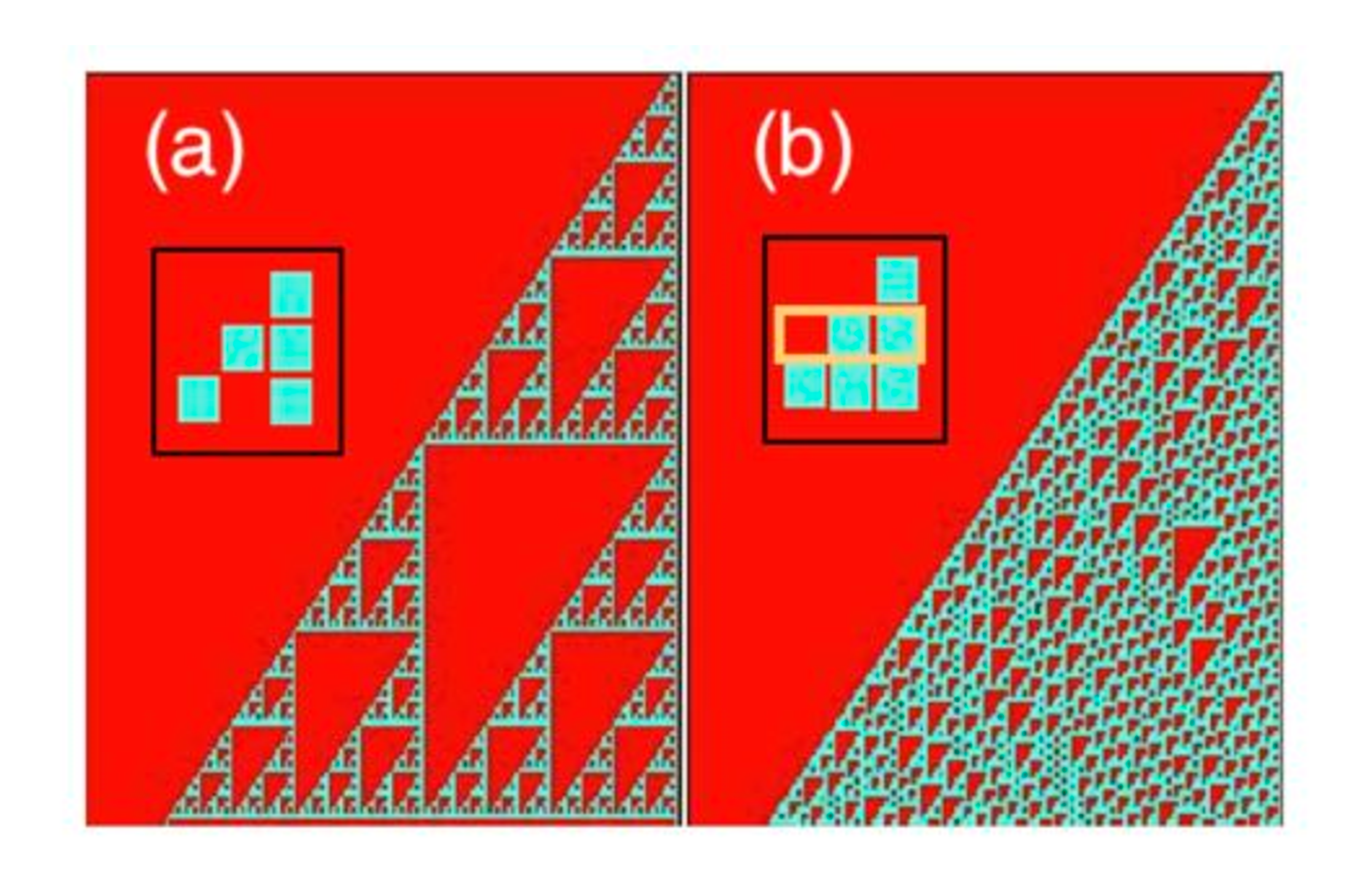}
\caption{(Color online) Spatiotemporal evolutions of: (a) the rule Eq. (\ref{wunder2}) ($p=2$); (b) Wolfram's rule 110, Eq. (\ref{Wolfram}) derived from the rule in (a) through a process of symmetry breaking illustrated in the inset (see main text). Time flows from top to bottom and the spatial dimension $i$ increases to the left in each panel.} \label{symbreak}
\end{figure}  

The most simple form of Eq. (\ref{wunder}) is $h=1$ (excluding the trivial one with $h=0$), and it is called henceforth a \emph{Pascal rule}. Such rule can be equivalently written as 
\begin{equation}
x_{t+1}^{i}=\sum_{s=0}^{\rho(p-1)}\sigma_{s}\mathcal{B}\left(s-\sum_{k=-r}^{l}x_{t}^{i+k},\frac{1}{2}\right) \label{CATnue}
\end{equation}
with $\sigma_{s}=0+_{p}s \equiv s \mod p$ \cite{morales2} ($+_{p}$ denotes addition mod $p$). $\mathcal{B}(y,\epsilon)=\frac{1}{2}\left(\frac{y+\epsilon}{|y+\epsilon|}-\frac{y-\epsilon}{|y-\epsilon|}\right) $ is the boxcar function of the two real valued quantities $y$ and $\epsilon$ \cite{morales1, morales2}. It returns '1' if $-\epsilon < y < \epsilon$ and '0' otherwise. Pascal rules are \emph{predictable}. The equation for the trajectory is given by the convolution theorem on the initial condition weighted by multinomial coefficients on certain planar sections of Pascal hyperpyramids \cite{Bondarenko}. When $\rho=2$ (we take $l=0$ and $r=1$) we have
\begin{equation}
x_{t+1}^{i}=x_{t}^{i}+_{p}x_{t}^{i-1} \label{wunder2}
\end{equation}  
and the solution for the trajectory is
\begin{equation}
x_{t}^{i}=\sum_{k=0}^{t}\binom{t}{k}x_{0}^{i-t+k} \mod p \label{solwunder2}
\end{equation}  
as can be shown by induction. For an initial condition with a single site with value '1' at $i=0$, surrounded by '0's  the spatiotemporal evolution yields the Pascal triangle modulo $p$, $x_{t}^{i}=\binom{t}{i} \mod p$. When $p=2$, it becomes the Sierpinsky triangle \cite{WolframM, Martin}, as shown in Fig. \ref{symbreak}a.

\section{Origin of complexity in cellular automata}
\label{breaking}

From (multi)fractal analysis and number theory, shortcuts can always be found, at least qualitatively, to predict the behavior of rules invariant under addition modulo $p$. Such shortcuts \emph{do not exist} for CA rules with Class 4 behavior, which \emph{necessarily} must then break this symmetry. 

\begin{figure}
\begin{center}
\includegraphics[width=0.53\textwidth]{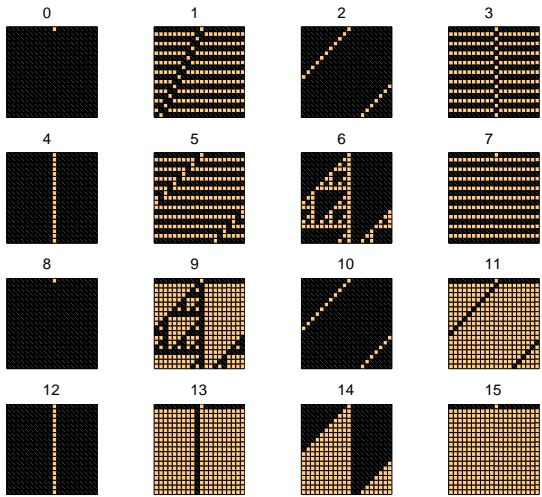}
\end{center}
\caption{(Color online) Spatiotemporal evolutions of the 16 rules with $p=2$, $r=1$, $l=0$ (range $\rho=2$). These rules constitute the 16 operators of boolean logic \cite{mathphys}. Rule labelled '6' in  (and its class equivalent under global complementation '9') are the only Class 3 rules among the 16 rules, being as well the only Pascal rules in the set. All other rules are Class 1 or 2. Time flows from top to bottom. A ring of 20 sites with periodic boundary conditions is considered.} \label{the16}
\end{figure}

We describe now a simple mechanism in which Class 4 CA rules do that. Let us consider the most simple CA rules with $p=2$ and $\rho=1$ and $2$. The 4 rules for $\rho=1$ and the 16 rules for $\rho=2$ (see Fig. \ref{the16}) are all easily predictable and their trajectory can be found by the inductive method \cite{morales1}. The most complex rule with $\rho=2$ (we take $r=1$, $l=0$: the same results hold for the 16 rules with $r=0$, $l=1$) is also the only Pascal rule, besides its class equivalent under global complementation, and has the form $x_{t+1}^{i}=x_{t}^{i}+_{2}x_{t}^{i-1}$.  These two rules are the only ones with Class 3 behavior among the 16 ones (all others are Class 1 or 2). \emph{Invariance  under addition modulo 2 singles out the most complex yet regular and predictable rules with $p=2$ and $\rho=2$}. With $p=2$, we need $\rho=3$ to find rules more complex than these and \emph{the symmetry under addition modulo 2 must be broken when adding the new degree of freedom}.  The symmetry breaking cannot be arbitrary: in order to at least keep the complexity already reached, we must demand that, for a given value of the added degree of freedom, the original Pascal rule behavior is regained. For the other value, this behavior should be distorted enough to introduce defects in the nested structure. A way of achieving this for rules with $\rho=3$ and $p=2$ is as sketched in Fig.\ref{symbreak}b: a term is added switching the output of the configuration '011' from '0' in the original Pascal rule to '1'. In this way, the borders of the triangle are not altered, the output of configuration '011' differs from the one of configuration '000' (breaking the symmetry upon addition modulo 2) and the Pascal rule behavior is regained when the site added to the left has value '1'. Since '011' equals 3 in base 2, this symmetry breaking operation can be introduced as
\begin{eqnarray}
x_{t+1}^{i}&=&x_{t}^{i}+_{2}x_{t}^{i-1}+_{2}\mathcal{B}\left(3-\sum_{k=-1}^{1}2^{k+1}x_{t}^{i+k},\frac{1}{2}\right)  \label{Wolfram}
\end{eqnarray}
The distribution of '0's within the triangle is now extremely irregular yet highly correlated: complexity arises from the interaction of the nested structure and the defects introduced through the symmetry breaking. For arbitrary initial conditions, the behavior is then also expected to display such a high complexity. In fact, the resulting rule, Eq. (\ref{Wolfram}), is Wolfram's 110 rule \cite{Wolfram5} (see Fig.\ref{symbreak}b), able to perform universal computation \cite{Wolfram1, Cook} and paradigmatic example of Class 4 behavior. The prescription to derive Class 4 CA rules is thus as follows: 1) consider Eq. (\ref{wunder}) for given values of $h$, $l$, $r$ and $p$; 2) From the nested structure obtained from the rule employing a simple initial condition, find which configurations output within its borders; 3) By introducing appropriate $\mathcal{B}$-terms, switch the output of any of these configurations (by adding new degrees of freedom if needed) such that for certain values of the latter Eq. (\ref{wunder}) holds, but defects are also introduced for other values, breaking the symmetry under addition modulo $p$ of the original rule. The resulting rule exhibits Class 4 behavior for arbitrary initial conditions. 

\begin{figure}
\begin{center}
\includegraphics[width=0.3\textwidth, angle=270]{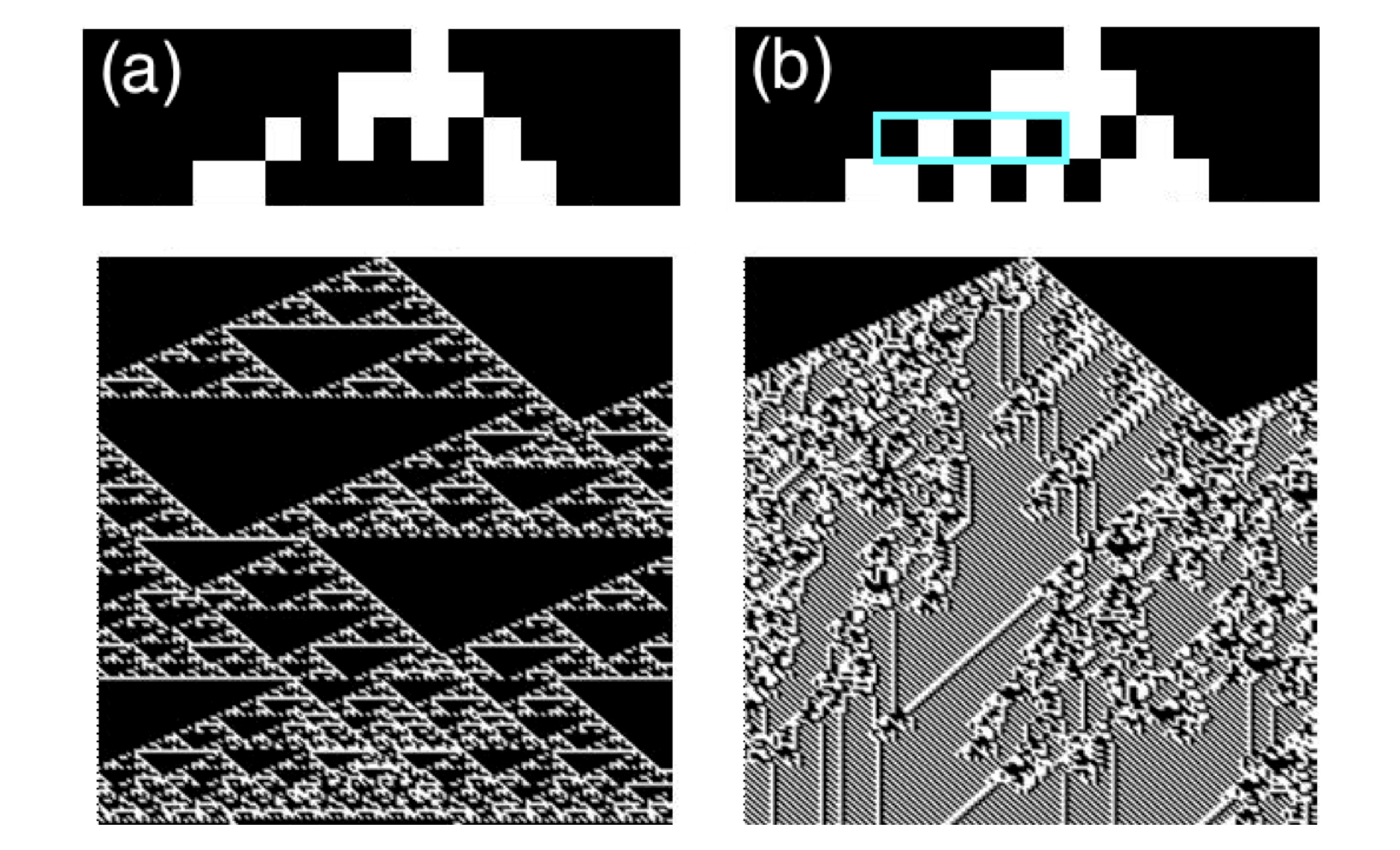}
\end{center}
\caption{(Color online) Spatiotemporal evolutions of (a) the Pascal rule given by Eq. (\ref{ppa} symmetric under addition modulo 2 and (b) the symmetry breaking rule in Eq. (\ref{ppb}). The detail shows the only configuration '01010' that is tuned compared to the symmetric Pascal rule (when the additional degree of freedom is added) in order to break the symmetry. Time flows from top to bottom. $i$ increases to the left. Shown is a window where $t \in [0,150]$ and $i\in [1, 150]$.} \label{symbreak2}
\end{figure}

A further nontrivial example of a rule constructed in this way is shown in Fig. \ref{symbreak2}. We consider the Pascal rule
\begin{equation}
x_{t+1}^{i}=x_{t}^{i+1}+_{2}x_{t}^{i}+_{2}x_{t}^{i-1}+_{2}x_{t}^{i-2} \label{ppa}
\end{equation}
with $(p=2, l=1, r=2, \rho=4)$. Its spatiotemporal evolution is shown in Fig. \ref{symbreak2}a. The above prescription  to find Class 4 rules can now be systematically followed. 1) We take the Pascal rule in Eq. (\ref{ppa}) as starting point. 2) We observe in Fig. \ref{symbreak2}a that, for example, configurations '01010','00011','01111' and '00010'  which correspond to numbers '10', '3', '31', '2' in the decimal system, fall within the triangle of evolution of the rule (see detail in Fig. \ref{symbreak2}a). 3) We can now switch the output of any of these configurations by introducing an additional degree of freedom. We introduce a term $\mathcal{B}\left(10-\sum_{k=-2}^{2}2^{k+2}x_{t}^{i+k},\frac{1}{2}\right)$ to switch the output of the configuration '01010' from '0' in the original Pascal rule to '1'. The resulting rule has map 
\begin{eqnarray}
x_{t+1}^{i}&=&x_{t}^{i+1}+_{2}x_{t}^{i}+_{2}x_{t}^{i-1}+_{2}x_{t}^{i-2} \nonumber \\
&&+_{2}\mathcal{B}\left(10-\sum_{k=-2}^{2}2^{k+2}x_{t}^{i+k},\frac{1}{2}\right) \label{ppb}
\end{eqnarray}
and exhibits Class 4 behavior (see (b) on the right, where the spatiotemporal evolution of this rule is shown) displaying gliders and glider guns (necessary elements to emulate logical gates and have universal computation). Finding this rule by brute force computation would require millions of simulations.

\section{Conditional predictability}
\label{conditional}

The symmetry under addition modulo $p$ can be broken in subtle ways leading to the formation of rare defects. The latter might be extremely difficult to find but they may have a decisive long-term impact in the global dynamics. We take $N_{s} \to \infty$ and consider $p$ a prime number and $m \in [0,p-1]$ an integer. The following rule breaks the symmetry under addition modulo $p$ of the Pascal rule in Eq. (\ref{wunder2}) 
\begin{equation}
x_{t+1}^{i}=x_{t}^{i}+_{p}x_{t}^{i-1}+_{p}m\mathcal{B}\left(1+p \uparrow\uparrow 3-
\sum_{k=0}^{p\uparrow\uparrow 2}p^{k}x_{t}^{i+k},\frac{1}{2}\right) \label{monster}
\end{equation}
Here Knuth's up-arrow notation \cite{Knuth} $p \uparrow\uparrow 3 \equiv p^{p^{p}}$ and $p \uparrow\uparrow 2 \equiv p^{p}$ has been used. For small $p$ these numbers are already enormous (e. g. $7 \uparrow\uparrow 2 = 823543$). The spatiotemporal evolution of Eq. (\ref{monster}) coincides on typical scales with Eq. (\ref{wunder2}) with trajectory given by Eq. (\ref{solwunder2}). However, although the defects caused by the $\mathcal{B}$-term are rare, they may happen. Indeed, we prove that if the initial condition is a single seed with value 1 at site $i=0$ a defect with value $m+_{p}1$ is introduced there at $p \uparrow\uparrow 2+1$ time steps. First note that from this initial condition it is not possible to have any defect before, since the highest non-zero power of $p$ within the sum in the $\mathcal{B}$-term in Eq. (\ref{monster}) is lower than $p \uparrow\uparrow 3$ and the $\mathcal{B}$-term outputs then '0'. The trajectory until $t=p \uparrow\uparrow 2$ is thus given by $x_{t}^{i}=\binom{t}{i} \mod p$, and at site $i=0$ the output will be $x_{t}^{0}=1$. We have, however 
\begin{eqnarray}
\sum_{k=0}^{p \uparrow\uparrow 2}p^{k}x_{p \uparrow\uparrow 2}^{k}&=&\sum_{k=0}^{p \uparrow\uparrow 2}p^{k}\left[\binom{p \uparrow\uparrow 2}{k} \mod p \right]\nonumber \\ &=& p^{p \uparrow\uparrow 2}+1=p \uparrow\uparrow 3+1 \label{jurl}
\end{eqnarray}
where we have used that 
\begin{equation}
\binom{p \uparrow\uparrow 2}{k} \mod p = 0 
\end{equation}
for $1 \le k \le p \uparrow\uparrow 2-1$. This result that can be proved by using Lucas' correspondence theorem \cite{Fine} which states that, for non-negative integers $m$ and $n$, and a prime $p$, the following  holds:
\begin{equation}
    \binom{m}{n}=\prod_{i=0}^k\binom{m_i}{n_i} \mod p 
\end{equation}
where
\begin{equation}
m=m_kp^k+m_{k-1}p^{k-1}+\cdots +m_1p+m_0 
\end{equation}
and
\begin{equation}
n=n_kp^k+n_{k-1}p^{k-1}+\cdots +n_1p+n_0 
\end{equation}
are the base $p$ expansions of $m$ and $n$. 

Therefore at time $t=p \uparrow\uparrow 2+1$ in site $i=0$ we have, from Eqs. (\ref{monster}) and (\ref{jurl}), 
\begin{eqnarray}
x_{p \uparrow\uparrow 2+1}^{0}&=&x_{p \uparrow\uparrow 2}^{0}+_{p}x_{p \uparrow\uparrow 2}^{-1} \nonumber \\
&&+_{p}\ m\mathcal{B}\left(1+p \uparrow\uparrow 3-\sum_{k=0}^{p\uparrow\uparrow 2}p^{k}x_{p \uparrow\uparrow 2}^{k},\frac{1}{2}\right) \nonumber \\
&=&1+_{p}m
\end{eqnarray}
as we wanted to prove. The trajectory until the next defect appears will then be given by Eq. (\ref{solwunder2}) but by replacing all $x_{0}^{i}$ by 
\begin{eqnarray}
x_{p \uparrow\uparrow 2+1}^{i}&=&\mathcal{B}\left(i-p \uparrow\uparrow 2-\frac{1}{2},1\right)+\mathcal{B}\left(i-1,\frac{1}{2}\right)
\nonumber \\
&&+\left(1+_{p}m\right)\mathcal{B}\left(i,\frac{1}{2}\right) 
\end{eqnarray}
 and $t$ by $t-p \uparrow\uparrow 2-1$, i.e. for $t \ge p \uparrow\uparrow 2+1$ we would thus have  
 \begin{equation}
 x_{t}^{i}=\sum_{k=0}^{t-p \uparrow\uparrow 2-1}\binom{t-p \uparrow\uparrow 2-1}{k}x_{p \uparrow\uparrow 2+1}^{i-t+p \uparrow\uparrow 2+1+k} \mod p
 \end{equation}
This means that we can know the whole trajectory piecewise explicitly only in rather special cases where we can also figure out when defects do happen. For arbitrary initial conditions, the trajectory must be checked for configurations yielding defects at each time. This is equivalent to simply running Eq. (\ref{monster}) with a computer: \emph{there is in general no shortcut for the trajectory of this CA}. This  CA is Class 4 (in huge time and spatial scales) but looks like regular Class 3 on shorter scales.  

In summary, there exists a subset of complex deterministic CA for which: 1) We  have an explicit expression for their trajectory which seems mostly valid; 2) We can know which kind of defects may appear and even know how they would influence the local and global dynamics in case they appear; however 3) for an arbitrary random initial condition we $cannot$ know if defects will indeed appear (if the system size is $N_{s} >> p\uparrow \uparrow 2$ in the example above, it is virtually $impossible$ to know the topology of all basins of attraction for all configurations in phase space containing the symbolic chain that introduces the defect); 4) as a consequence of 3) the explicit expression for the trajectory of the system can only be \emph{conditionally} accepted. We call \emph{conditionally predictable CA} those which meet all above observations. Note that if we consider an arbitrary natural number $n>2$, all systems described by 
\begin{equation}
x_{t+1}^{i}=x_{t}^{i}+_{p}x_{t}^{i-1}+_{p}m\mathcal{B}\left(1+p \uparrow\uparrow n-
\sum_{k=0}^{p\uparrow\uparrow (n-1)}p^{k}x_{t}^{i+k},\frac{1}{2}\right) 
\end{equation}
are conditionally predictable for $p$ sufficiently large, independently of parameters $m$ and $n$.

\section{Discrete Chimera-like states} 
\label{chimeras}

\begin{figure}
\includegraphics[width=0.52\textwidth, angle=270]{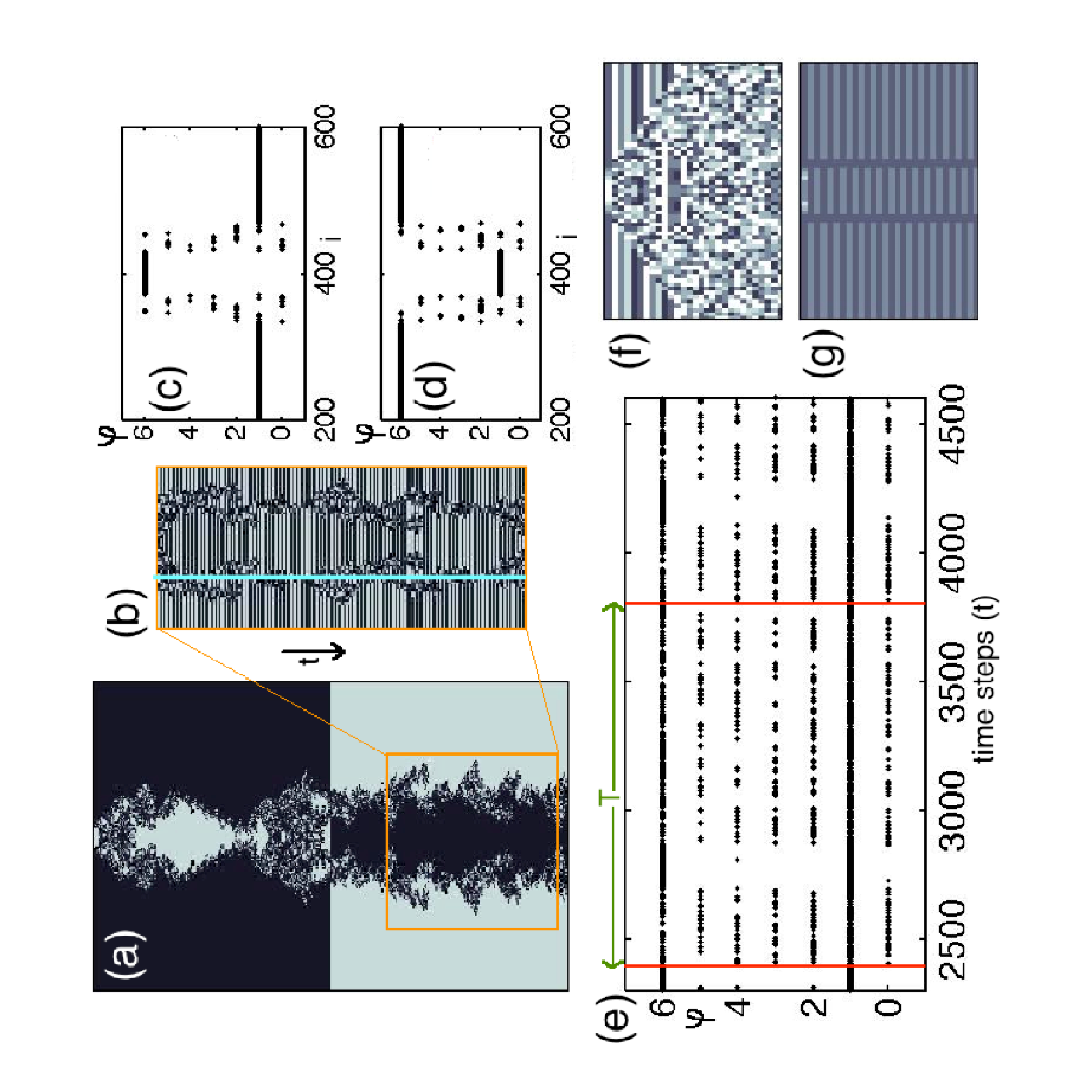}
\caption{(Color online) (a) Spatiotemporal evolution of Eq. (\ref{CATphase}) with $p=7$, $l=r=2$ ($\rho=5$) and $C=16$ for an initial condition consisting of $38$ sites with value '1' surrounded by '0's on a ring of $N_{s}=800$ sites and for a time window $t \in [0,4000]$ time steps, with time flowing from top to bottom (note that the homogeneous background black and white regions contain actually 2000 oscillations between black and white that cannot be displayed in (a)). (b) Detail of the stable incoherent structure. (c) Snapshot (detail) of the spatial phase distribution for $t=2000$ and (d) $t=2001$ time steps. (e) Local time series for an oscillator in the incoherent region corresponding to the cut in (b) where a duration spanning the recurrence time $T=1408$ of the full incoherent structure is indicated. (f)-(g) Spatiotemporal evolutions (detail) for  $C=30$ and  $C=6$ respectively (other parameter values as in (a)).} \label{chimer} 
\end{figure}

Chimera states arise in an ensemble of non-locally coupled identical nonlinear oscillators when synchrony and incoherence coexist in separated stable  domains \cite{Kuramoto, Abrams, Scholl1, Scholl2, Showalter}. We construct now a CA model for discrete chimera-like states with a number $p$ of states for the phase $\varphi$ (i.e. $\varphi \in [0,p-1]$ integer) given in $2\pi/p$ units.  We follow the prescription to find Class 4 CA rules: 1) We take Eq. (\ref{CATnue}) as starting point. 2) We observe that configurations with sum-over-neighborhood values larger than $p-2$ output within the nested structure. 3) We design the symmetry breaking so that all configurations with sum-over-neighborhood values exceeding a threshold value $C \in [p-1,\rho(p-1)]$ output a constant value $a \in [0,p-1]$. For $C$ low, this may eventually cause the dynamics to collapse into uniform states. For $C$ large, the behavior is incoherent. For $C$ intermediate the coexistence of synchrony and incoherence is expected. We note that adding modulo $p$ an integer constant $b \in [0,p-1]$ to the r.h.s. of Eq. (\ref{CATnue}) introduces a collective rotation without affecting the symmetry of the Pascal rule. In the following, we take $b=1$ and $a=0$. From all above, the CA model for the phase dynamics reads
 \begin{equation}
\varphi_{t+1}^{i}=\sum_{s=0}^{\rho(p-1)}\alpha_{s}\mathcal{B}\left(s-\sum_{k=-r}^{l}\varphi_{t}^{i+k},\frac{1}{2}\right) \label{CATphase}
\end{equation}
with
\begin{eqnarray}
\alpha_{s}&=&1+_{p}
\left[s \mathcal{B}\left(\frac{C-1}{2}-s, \frac{C}{2}\right)+\mathcal{B}\left(C-s+\frac{1}{2},1 \right)\right] \nonumber \label{alphy} \\
\end{eqnarray}
The first $\mathcal{B}$-term in Eq. (\ref{alphy}) controls the number of configurations that follow the Pascal rule. For $s$ larger or equal than $C$ the term outputs '0', breaking the symmetry.  The second $\mathcal{B}$-term outputs '1' only for $s=C$ and $s=C+1$ and '0' for any other $s > C+1$. It provides a further symmetry breaking although it is essentially unnecessary: its only effect is a longer recurrence time of the incoherent region.

We take $p=7$, $l=r=2$ ($\rho=5$) in Eq. (\ref{CATphase}). Then $C \in [6,30]$.  In Fig. \ref{chimer}a the spatiotemporal evolution of Eq. \ref{CATphase} for $C=16$ is shown. After a transient of ca. 1900 time steps, the system displays a stable coexistence of synchrony and incoherence. In Fig. \ref{chimer}b a detail of this state is shown. Uniformly oscillating domains with phase difference $\Delta \phi =5$ coexist with two incoherent structures that mirror each other (the CA dynamics is invariant under reflection and the initial condition is symmetric). This is also clear from Fig. \ref{chimer}c and d, where snapshots of the phase distributions vs. position at times $t=2000$ and $2001$ are shown. This coexistence is \emph{eternal}: the incoherent region then necessarily repeats itself after a recurrence time $T$. Were it $ergodic$ and $\delta$ sites thick, we would have $T=T_{e}=p^{\delta}$. In general $T << T_{e}$ since the CA behavior is highly correlated. The time series of $\varphi$ for one site in the incoherent region exhibits an erratic jumping between its seven possible values (as shown in Fig. \ref{chimer}e) revealing $T=1408$ time steps, which is \emph{three orders of magnitude longer than the period of the uniformly oscillating domains}. This would correspond to $\delta=\ln T/\ln p \approx 3.726$, were the incoherent region ergodic, contrasting with the observed value $\delta \approx 90$. In Fig. \ref{chimer}f and \ref{chimer}g the spatiotemporal evolutions are plotted for $C=30$ and $C=6$, respectively. For $16 < C < 30$ all site values tend to behave incoherently (as in Fig. \ref{chimer}f) while for $C< 16$ all oscillators are fully synchronized after a certain transient arranging into phase clusters as in Fig. \ref{chimer}g. A detailed analysis of chimera states as a function of $p$, $\rho$ and $C$ will be presented elsewhere. 

In this article, a prescription involving the breaking of the symmetry under addition modulo $p$ has been shown to originate complexity out of simple predictable rules. This idea has been used to reveal \emph{conditionally predictable} cellular automata, whose trajectory can be piecewise known with certainty only if the positions of possible (rare) defects are also known. The mechanism for complexity has been fruitfully used to construct a model for discrete chimera-like states. The incoherent region has been shown to have a recurrence time that is three orders of magnitude longer than the period of the synchronous domains. A most interesting feature of this kind, at the interface between dynamics and statistical mechanics, cannot be captured with previous existing models where a real-valued phase $\varphi \in [0,2\pi]$ \cite{Kuramoto, Abrams, Scholl1}) is considered. Although having a quantized oscillator's phase may seem unnatural, the limit $p\to \infty$ can be taken within the theory (details will be discussed elsewhere), and studying the dynamics behind the discrete case is rewarding in itself. Also, there are certain (probabilistic) models in quantum Physics where only just a finite number of values for the quantum phase are considered (as the Feynman checkerboard model \cite{Hibbs}, where the phase only takes as values the 4th roots of unity). 

I thank Katharina Krischer and Lennart Schmidt for conversations. Support from the Technische Universit\"at M\"unchen - Institute for Advanced Study, funded by the German Excellence Initiative, is gratefully acknowledged.

\end{document}